\documentclass[12pt]{iopart}
\usepackage{graphicx}
\begin{document}

\title[A study of resonances in high intensity beams using a linear Paul trap]{A study of coherent and incoherent resonances in high intensity beams using a linear Paul trap}

\author{L K Martin$^1$, S Machida$^2$, D J Kelliher$^2$ and S L Sheehy$^1$}

\address{$^1$ Physics Department, University of Oxford, Keble Road, Oxford, OX1 3RH, UK}
\address{$^2$ STFC, Rutherford Appleton Laboratory, Harwell Oxford, Didcot, OX11 0QX, UK}
\ead{lucy.martin@physics.ox.ac.uk}

\begin{abstract}
In this paper we present the first quantitative measurement of the change in frequency (tune) with intensity of four transverse resonances in a high intensity Gaussian beam. Due to the non-linear space charge forces present in high intensity beams, particle motion cannot be analytically described. Instead we use the Simulator of Particle Orbit Dynamics (S-POD) and the Intense Beam Experiment (IBEX),  two linear Paul traps, to replicate the system experimentally.
In high intensity beams a coherent resonant response to both space charge and external field driven perturbations is possible, these coherent resonances are excited at a tune that differs by a factor $C_{m}$ from that of the incoherent resonance. By increasing the number of ions stored in the linear Paul trap and studying the location of four different resonances we extract provisional values describing the change in tune of the resonance with intensity. These values are then compared to the $C_{m}$ factors for coherent resonances. We find that the $C_{m}$ factors do not accurately predict the location of resonances in high intensity Gaussian beams. Further insight into the experiment was gained through simulation using Warp, a particle-in-cell code.
\end{abstract}

%
\vspace{2pc}

\submitto{\NJP}
%
%
%
\section{\label{sec:Introduction}Introduction}

High intensity particle accelerators are vital in many applications, from spallation neutron sources to the transmutation of nuclear waste. However, in high intensity machines the Coulomb interactions between particles, known as space charge forces, cannot be neglected. The magnitude of these space charge forces impose the ultimate limit on beam intensity in recirculating accelerators.

Coulomb forces between charged particles in the beam of an accelerator are repulsive, acting as a lens which defocuses in both transverse planes \cite{Schindl}. As the intensity of a beam is increased the strength of this defocusing becomes greater, resulting in a shift in the oscillation frequency, called the tune, of circulating particles in high intensity beams.

As the transverse beam distribution in an accelerator does not generally lead to linear defocusing (except when a KV distribution is assumed \cite{Kapchinski1959}), space charge forces are a nonlinear perturbation to the linear focusing force produced by the magnets. Each particle therefore experiences a different tune shift, resulting in an overall tune spread within the beam. A greater spread in  tune makes it more likely that some particles will have a tune lying within the stopband of a resonance, where they can be resonantly excited, leading to increased beam loss.  

It is known that a single particle experiences a resonance when its tune agrees with the frequency of a periodic driving force. This is described by the resonant condition

\begin{equation}
\label{eq:IncoherentResonance}
Q_{\rm 0} = \frac{n}{m},
\end{equation}

\noindent
where $Q_{\rm 0}$ represents the bare tune of the particle (not including any tune shift from space charge forces) and $n$ and $m$ the harmonic and order of the perturbing driving force respectively, so that $m=1$ corresponds to a dipole perturbation, $m=2$ to a quadrupole perturbation and so on. 

Including the tune shift due to space charge forces, $\Delta Q$, equation (\ref{eq:IncoherentResonance}) becomes

\begin{equation}
\label{eq:IncoherentResonance2}
Q_{\rm 0} + \Delta Q= \frac{n}{m},
\end{equation}

\noindent
where $ \Delta Q$ is negative, known as the tune depression.

In the past accelerators were designed so that the incoherent tunes of the particles in the beam avoided this resonant condition \cite{Morin1962}. However, the increased space charge forces at high intensity can result in the beam responding as a whole to a perturbation. This can lead to a collective motion of the beam, known as coherent oscillation \cite{Baartman1998, Lapostolle1963, Smith1963}.

It is the coherent oscillation of a beam that leads to the excitation of the greatest number of particles and therefore the greatest beam loss \cite{Sacherer1968}. In the case of coherent oscillation a self-consistent solution to the equations of motion of the beam is necessary to describe the resonant response. For a quadrupole ($m$=2) coherent oscillation this is done by constructing a set of envelope equations to describe the beam motion along the trajectory $s$, where $s$ the path length in the beam direction \cite{Sacherer1968},

\begin{eqnarray}
\label{eq:EnvelopeEquations}
\frac{\rmd^{2}a}{\rmd s^{2}} + K_{\rm a}(s)a - \frac{\epsilon_{\rm a}^{2}}{a^{3}} - \frac{K_{\rm sc}}{2(a + b)} = 0, \\
\frac{\rmd^{2}b}{\rmd s^{2}} + K_{\rm b}(s)b - \frac{\epsilon_{\rm b}^{2}}{b^{3}} - \frac{K_{\rm sc}}{2(a + b)} = 0.
\end{eqnarray}

In these envelope equations $a$ and $b$ are the root mean square (rms) sizes of the beam in the transverse directions and $\epsilon_{a}$ and $\epsilon_{b}$ the rms emittance. The periodic focusing is described by $K(s)$ and the space charge defocusing is given by

\begin{equation}
K_{\rm sc} = \frac{N q^{2}}{2 \pi \epsilon_{\rm 0} m \gamma^{3} v^{2}},
\end{equation}

\noindent
also known as the perveance. Here $q$ is the charge of the particle, $m$ its rest mass, $N$ the line density of the beam, $\gamma$ the Lorentz factor and $v$ the beam velocity. 

By applying the smooth approximation and solving the envelope equations the new resonant condition is found to be

\begin{equation}
\label{eq:CoherentResonance2}
Q_{\rm0} + C_{2}\Delta Q = \frac{n}{2}.
\end{equation}

\noindent
where $C_{2} = \frac{3}{4}$ when the tunes in both transverse directions are equal \cite{Sacherer1968}. The condition can be generalised to higher order $m$ by the linearised Vlasov-Poisson equation, giving

\begin{equation}
\label{eq:CoherentResonance}
Q_{\rm 0} + C_{m}\Delta Q = \frac{n}{m}.
\end{equation}

The $C_{m}$ factor now present in the resonant condition depends on the mode of coherent oscillation that the beam is experiencing \footnote{$C_{mk}$ is also commonly used, where $m$ denotes the azimuthal mode of oscillation and $k$ the radial mode. Here $k$ is dropped as $m=k$ is assumed.}.

It has been analytically shown that the $C_{m}$ factors all lie in the range $0<C_{m}<1$, with the special case of $C_{1} = 0$ for a dipole perturbation \cite{Sacherer1968}. As the order of the coherent oscillation increases so does the $C_{m}$ factor, so that $C_{1}<C_{2}<C_{3}...$ etc. As the $C_{m}$ factors are less than $1$ this gives a small increase in the high intensity beam limit, meaning that the intensity of the beam can be increased further before the coherent resonant condition is met. 

This work was later extended by Okamoto and Yokoya in reference \cite{Okamoto2002}, where no smooth approximation is applied. The theory presented in \cite{Okamoto2002} shows that, when alternating gradient focusing is considered, space charge driven resonances emerge at twice the density in tune space, according to the resonant condition

\begin{equation}
\label{eq:SpacechargeResonance}
Q_{\rm 0} + C_{m}\Delta Q = \frac{1}{2}\big(\frac{n}{m}\big).
\end{equation}

The growth rate and stop-band width of these space charge driven resonances were found to be proportional to the perveance, and therefore the intensity of the beam. Such resonances, also referred to as structure space charge instabilities or coherent parametric instabilities \cite{Hofmann2017}, are therefore neglected at low intensities.

Analytical values of $C_{m}$ have been calculated assuming a Kapchinski-Vladimirskij (KV) distribution, where the space charge forces provide linear defocusing in both planes. These values can then be used to describe different particle distributions assuming rms equivalent beams. These $C_{m}$ values are presented in Table \ref{table:TheoreticalCm} \cite{Sacherer1968}.

\begin{table} 
\caption{\label{table:TheoreticalCm}Table showing the analytically calculated $C_{m}$ values for a 2D KV beam \cite{Sacherer1968}.}
\begin{indented}
\lineup
\item[]\begin{tabular}{@{}*{5}{l}}
\br
Order of coherent resonance ($m$)& 1  & 2   &  3  & 4\\
\mr
$C_{m}$ & 0 & $\frac{3}{4}$ &  $\frac{11}{12}$ &  $\frac{31}{32}$ \\
\br
\end{tabular}
\end{indented}
\end{table}

Additionally, the transverse distribution of the beam itself is thought to determine whether a beam is excited coherently. Hofmann, in reference \cite{Hofmann2017-2}, shows that KV distributions can be excited coherently by higher order modes, but that for more realistic beam models, such as waterbag or Gaussian distributions, the spread in particle tunes may lead to Landau damping of coherent modes higher than $m=2$. For these higher order modes the significant overlap between the incoherent tunes in the distribution and the tune of the coherent resonance can lead to energy transfer from the coherent excitation of the distribution to the incoherent excitation of particles within the beam, so that coherent oscillations never form \cite{Landau1946}. 

The resonances occurring at a cell tune of $\frac{1}{4}$ have been studied in simulation a number of times  \cite{Hofmann2015, Jeon2009, Li2014}. Simulations show clearly the competition between the fourth order incoherent excitation of the beam and the second order coherent resonance at this tune. 

Accurate knowledge of the location of resonances in high intensity beams is desirable so that accelerators can be designed to avoid such resonances, minimising beam loss in high intensity machines. However, due to the non-linear space charge forces present in high intensity beams even particle motion in an accelerator with purely linear elements cannot be analytically described, it can only be approximated. Further complication is introduced by the emittance change due to particle loss near a resonance, which is not accounted for in the analytical calculation of the $C_{m}$ factors; the potential for Landau damping of coherent excitation; and the competition between resonances of different orders at the same tune.

Simulating a high intensity accelerator numerically requires a number of simplifying assumptions and is computationally intensive, making the study of high intensity resonances challenging. Instead, a Linear Paul Trap (LPT) can be used to recreate the alternating gradient (FODO) lattice of an accelerator experimentally. A LPT can study high intensity beam dynamics in such lattices by trapping a large number of argon ions, which have a transverse Gaussian distribution, in an electrical quadrupole potential. By varying the tune of the trapped ions, ion loss due to resonance can be studied.

The Simulator of Particle Orbit Dynamics (S-POD) at Hiroshima University, Japan, and the Intense Beam Experiment (IBEX) at the Rutherford Appleton Laboratory, UK, are LPTs designed for high intensity accelerator physics. A significant amount of previous work exists using S-POD to show the overlap in high intensity beams of external field driven resonances (satisfying the resonant condition in equation (\ref{eq:CoherentResonance})) and space charge driven resonances (meeting the resonant condition in equation (\ref{eq:SpacechargeResonance})) \cite{Ito2017}. 
Furthermore, the splitting between the location of the externally driven coherent dipole resonance and self driven coherent quadrupole resonance at integer resonances, due to the differing $C_{m}$ factors, has been shown qualitatively \cite{Moriya2016}. As expected, this splitting becomes more pronounced as intensity is increased.  

In reference \cite{Ohtsubo2010} Ohtusubo et al. applied a quadrupole error to the S-POD trap to excite resonances and ion loss was studied at three different intensities. They qualitatively draw the conclusion that the $C_{m}$ factors for $m = 2$, $m = 3$ and $m = 4$ ``appear to be less than unity'' at high intensities. However, the underlying assumptions of this study do not allow the result to be quantified. In this paper we study resonances in high intensity Gaussian beams over longer time scales and quantitatively for the first time, a significant advancement in the use of Paul ion traps to study high intensity effects in particle accelerators.

In this paper we build on this previous work to study resonances in high intensity Gaussian beams over longer time scales and quantitatively for the first time, a significant advancement in the use of Paul ion traps to study high intensity effects in particle accelerators. To start, we do not assume that resonances are either coherent or incoherent, we only assume that the resonance condition for maximal beam loss can be defined as

\begin{equation}
\label{eq:A}
Q_{\rm 0} + A_{m}\Delta Q = \frac{1}{2}\big(\frac{n}{m}\big).
\end{equation}

\noindent
We use the S-POD LPT to extract a value for $A_{m}$, which is of practical use in the design of high intensity accelerators.

We varied the cell tune (tune per focusing period) in the S-POD LPT over a wide range and studied the location of the resonances experienced at a range of intensities. We do not apply any external error fields to the trap, resonances are either driven by space charge forces or by small multipole fields from slight trap misalignments and in the high intensity regime in which this experiment is conducted space charge driven resonances dominate \cite{Ohtsubo2010} . From this data we extracted a numerical value for the location of four resonances of different orders at different ion numbers. By fitting to this we extract $A_{m}$, which we then compare to the $C_{m}$ factors predicted for purely coherent resonances. 

To understand the experiment further we simulated the setup in the particle-in-cell (PIC) code Warp. 

We first briefly describe the experimental setup of the S-POD and IBEX LPTs in section \ref{sec:Linear Paul Traps}. Our experimental procedures are shown in section \ref{sec:Experimental procedure and results}.  We then present our simulation of the system in section \ref{sec:Simulation}, as the results helped to guide the analysis of the experimental data. The analysis of the experimental data is described in section \ref{sec:Analysis}.

\section{Linear Paul Traps\label{sec:Linear Paul Traps}}

A LPT stores ions using an alternating voltage applied to four cylindrical rods, which creates an electrical quadrupole potential of the form 

\begin{equation}
\label{eq:potential}
U = \frac{V(t)}{2r_{\rm 0}^{2}}(x^{2} - y^{2}).
\end{equation}

\noindent
Here $r_{0}$ is the inscribed radius of the LPT rods, $x$ and $y$ are the usual transverse coordinates and $V(t)$ is the time dependent voltage applied to the rods.

The resultant LPT has a transverse Hamiltonian equivalent to an alternating gradient lattice in an accelerator, meaning that transverse particle motion in the two systems is the same \cite{Okamoto2002Trap, Davidson2000}.

The Hamiltonian for a LPT is expressed as

\begin{equation}
\label{eq:PaulHamiltonian}
H_{\rm Paul} = \frac{(p_{x}^{2} + p_{y}^{2})}{2} + \frac{1}{2}K_{\rm P}(\tau)(x^{2} - y^{2}) + \frac{q}{mc^{2}}\phi_{\rm sc},
\end{equation}

\noindent
where the focusing term is given by

\begin{equation}
\label{eq:PaulFocusing}
K_{\rm P}(\tau) = \frac{2qV(\tau)}{mc^2r_{0}^{2}}.
\end{equation}

\noindent
Here $p_{x}$ and $p_{y}$ are the particle momenta in the transverse directions, $q/m$ is the ion charge to mass ratio and $c$ the speed of light. The potential due to the space charge forces is denoted by $\phi_{\rm sc}$ and $\tau = ct$.

The Hamiltonian in equation (\ref{eq:PaulHamiltonian}) can be directly compared to the Hamiltonian of particle in a FODO cell of an accelerator, 

\begin{equation}
\label{eq:FODOHamiltonian}
H_{\rm beam} = \frac{(p_{x}^{2} + p_{y}^{2})}{2} + \frac{1}{2}K(s)(x^{2} - y^{2}) + \frac{q}{p_{\rm 0}\beta_{\rm 0}c\gamma_{\rm 0}^{2}}\phi,
\end{equation}

\noindent
where the focusing is now described by

\begin{equation}
\label{eq:FODOFocusing}
K(s) = -\frac{q}{p_{\rm 0}}\frac{\rmd B_{z}}{\rmd x} = -\frac{1}{B\rho}\frac{\rmd B_{z}}{\rmd x}.
\end{equation}

Equation (\ref{eq:FODOFocusing}) describes the focusing due to a quadrupole magnet, where $B\rho$ is the magnetic rigidity and $\rmd B_{z}/ \rmd x$ the quadrupole gradient. 

The third term on the right hand side of equation (\ref{eq:FODOHamiltonian}) (describing the space charge forces) now includes the Lorentz factor, $\gamma_{\rm 0}$, the ratio of the particle velocity to the speed of light, $\beta_{\rm 0}$ and the forward momentum of the particle, $p_{\rm 0}$. These terms are present as the interactions between relativistic line currents compete with static Coulomb interactions.

The LPT is a compact experimental system which allows a large number of cell tunes to be accessed by changing the amplitude of the voltage applied to the four confining rods. Equation (\ref{eq:PaulFocusing}) shows the effect of changing the voltage on the focusing term of the Hamiltonian. The number of ions stored can also be varied over a large range, allowing space charge effects to be investigated at different intensities. Additionally, a large ion loss does not damage a LPT or irradiate it, making it a useful tool in the study of beam loss at a resonance. It should be noted that dispersion due to momentum spread is not modelled in a LPT as the confining quadrupole is electric, however, such effects are not relevant to this study.

\subsection{\label{sec:Emittance and Temperature in a linear Paul trap}Emittance and Temperature in a linear Paul trap}

To understand the transverse dynamics of a LPT fully the emittance of the trapped ions must be known. The temperature and therefore the emittance in a Paul trap can be estimated from the envelope equations. 

Particle motion in a LPT is non-relativistic and so magnetic forces can be neglected. Considering only the horizontal direction and changing the independent variable from $s$ to $\tau$ through $s = \beta_{0}\tau$, equation (\ref{eq:EnvelopeEquations}) becomes

\begin{equation}
\label{eq:EnvelopePaul}
\frac{\rmd^{2}a}{\rmd \tau^{2}} + K_{\rm a}(\tau)a - \frac{\epsilon_{\rm a}^{2}}{a^{3}} - \frac{N_{\rm tot}r_{\rm p}}{2a} = 0,
\end{equation}

\noindent
in the non-relativistic approximation, where $N_{\rm tot}$ is the total line density of the ion cloud, $\epsilon_{\rm a}$ is the rms emittance and $r_{\rm p}$ is the classical particle radius of the trapped ions. 

Following the analysis in reference \cite{Okamoto2002Trap} the rms emittance can be expressed as

\begin{equation}
\label{eq:emit}
\epsilon_{\rm a} = a  \sqrt{\frac{K_{\rm B} T}{m c^{2}} },
\end{equation}

\noindent
where $T$ is the ion distribution temperature and $K_{\rm B}$ the Boltzmann constant.

Using this definition and assuming a stationary ion distribution ($\rmd ^{2}a/\rmd\tau^{2} = 0$) equation (\ref{eq:EnvelopePaul}) can be rearranged to give an expression for the rms size of the beam in terms of transverse temperature, 

\begin{equation}
\label{eq:size}
a = \frac{c}{\omega_{\rm q}}\sqrt{\frac{K_{\rm B} T}{m c^{2}} + \frac{N_{\rm tot}r_{\rm p}}{2}}.
\end{equation}

\noindent
Here $\omega_{\rm q}$ is the angular frequency of the betatron oscillations of the trapped ions. 

Rearranging this equation and defining the effective value of the incoherent phase advance as 

\begin{equation}
\label{eq:phaseAdvance}
\sigma^{2} = \sigma_{\rm 0}^{2} - \frac{N_{\rm tot} r_{\rm p}}{2} \bigg( \frac{\lambda}{a}\bigg)^{2},
\end{equation}

\noindent
where $\sigma_{\rm 0}$ is the bare phase advance and $\sigma$ is the shifted (depressed) phase advance, the following expression for transverse temperature in terms of tune depression can be derived, 

\begin{equation}
\label{tuneDepression}
T = \frac{N_{\rm tot} r_{\rm p} m c^{2}}{2 K_{\rm B}} \bigg( \frac{1}{1 - \eta^{2}} - 1 \bigg).
\end{equation}

\noindent
The tune depression is defined to be $\eta = \sigma/\sigma_{\rm 0}$.

\subsection{\label{sec:Structure of a Paul trap}Structure of a Paul trap}

The experimental setups of both the S-POD and IBEX LPTs have been described in detail in a number of previous publications \cite{Okamoto2002Trap, Davidson2000,Sheehy2017} and so will only be briefly outlined here.

\begin{figure}
\makebox[\columnwidth][c]{\includegraphics[width=0.7\columnwidth]{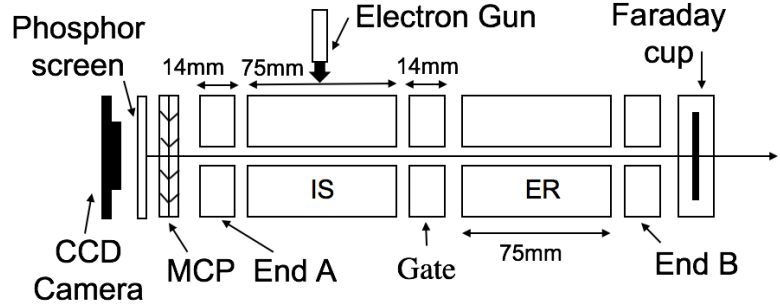}}
\caption{Diagram showing the layout of the S-POD LPT. Here the region labeled IS is the Interaction Section, where the ions are stored, and ER the Experimental Region. During this study the ER was used only during the extraction of ions. IBEX is identical in construction except that the ER section and gate are removed, allowing direct extraction onto the Faraday cup. Image adapted from \cite{Ito2015a}.}
\label{fig:SPODSetup}
\end{figure}

Figure \ref{fig:SPODSetup} shows a diagram of the trapping region in both S-POD and IBEX. The trapping region sits inside a larger vacuum vessel which maintains an Ultra High Vacuum (UHV) of $\sim 10^{-10}$ mbar. Before performing experiments, argon gas is introduced to the vessel through a leak valve. 

The argon gas is ionised by an electron gun directed into the space between the four cylindrical trapping rods. It is the $^{40}$Ar$^{+}$ ions created in this electron bombardment process that are confined by the rods and used in the high intensity beam experiments. 

\begin{figure}
\makebox[\columnwidth][c]{\includegraphics[width=0.4\columnwidth]{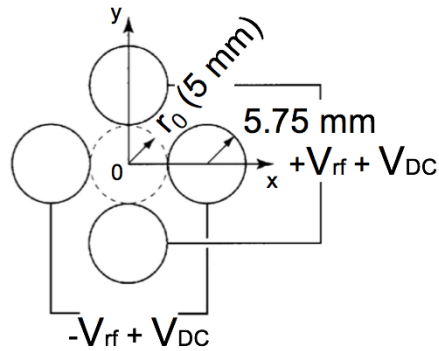}}
\caption{Diagram showing geometry of and the voltage applied to the confining rods in the S-POD and IBEX Paul traps. Image adapted from \cite{Okamoto2002Trap}.}
\label{fig:SPODRods}
\end{figure}

Ions are confined longitudinally using the endcaps, sections of shorter cylindrical rod not electrically connected to the main confining rods (End A, End B and Gate in figure \ref{fig:SPODSetup}). A positive DC offset is applied to these rods, added to the alternating voltage by an Arbitrary Wavefunction Generator (AWG) before amplification. This creates a longitudinal potential well, trapping the ions. As the confining rods are much longer than the end caps an almost rectangular potential well is produced, leading to an approximately homogeneous longitudinal ion distribution.

The trap confines the ions transversely using an alternating voltage applied to the four cylindrical rods, as shown in figure \ref{fig:SPODRods}. Ideally hyperbolic electrodes would be used to create the perfect quadrupole potential, however, cylindrical rods are much easier to machine accurately and give a very good approximation to a perfect quadrupole in the centre of the trap, with the addition of some small higher order modes \cite{Ohtsubo2010}. The transverse confining potential (equation (\ref{eq:potential})) is created by applying either a sinusoidal voltage or a step function to the rods. The rf trapping voltage is supplied using two synchronised channels from the AWG (created separately) with the rf components 180 degrees out of phase, which are then amplified. The rods are powered in pairs, with opposing rods electrically connected. The IBEX and S-POD systems are operated with a sinusoidal voltage of 1 MHz and the transverse focusing is controlled by varying the amplitude of the applied voltage. The cell tune ($Q$) in the trap, not including the negligible term due to the endcap voltage, is approximately 

\begin{equation}
\label{eg:tune}
Q \approx \frac{q V_{\rm rf}}{m \omega_{\rm rf}^{2} r_{\rm 0}^{2}},
\end{equation}

\noindent
where $V_{\rm rf}$ is the amplitude of the applied radio frequency voltage, $\omega_{\rm rf}$ is its angular frequency and $r_{\rm 0}$ the inscribed radius of the trap \cite{Okamoto2002}. 

After an experiment is performed, i.e. after the ions have been stored for a given time under certain conditions, the ions are extracted either onto a Faraday cup or a Multi Channel Plate (MCP). Both of these detectors are entirely destructive. To extract ions from the trap the DC component of one of the endcap voltages is switched off so that ions are no longer confined longitudinally. Throughout the experiment a positive DC potential offset is applied to the central rods, this ensures that ions are forced out of the trap when an endcap voltage is removed. 

The MCP contains a number of channels which amplify the signal through an electron cascade. The amplified electrons hit a phosphor screen, emitting light which is photographed by a CCD camera. The MCP can be used to give an integrated image of the ion distribution or, when calibrated, act as an ion counter. The Faraday cup on the other end of the trapping region is used as an ion counter.

\section{\label{sec:Experimental procedure and results}Experimental procedure and results}

\begin{figure*}
\makebox[\columnwidth][c]{\includegraphics[width=\columnwidth]{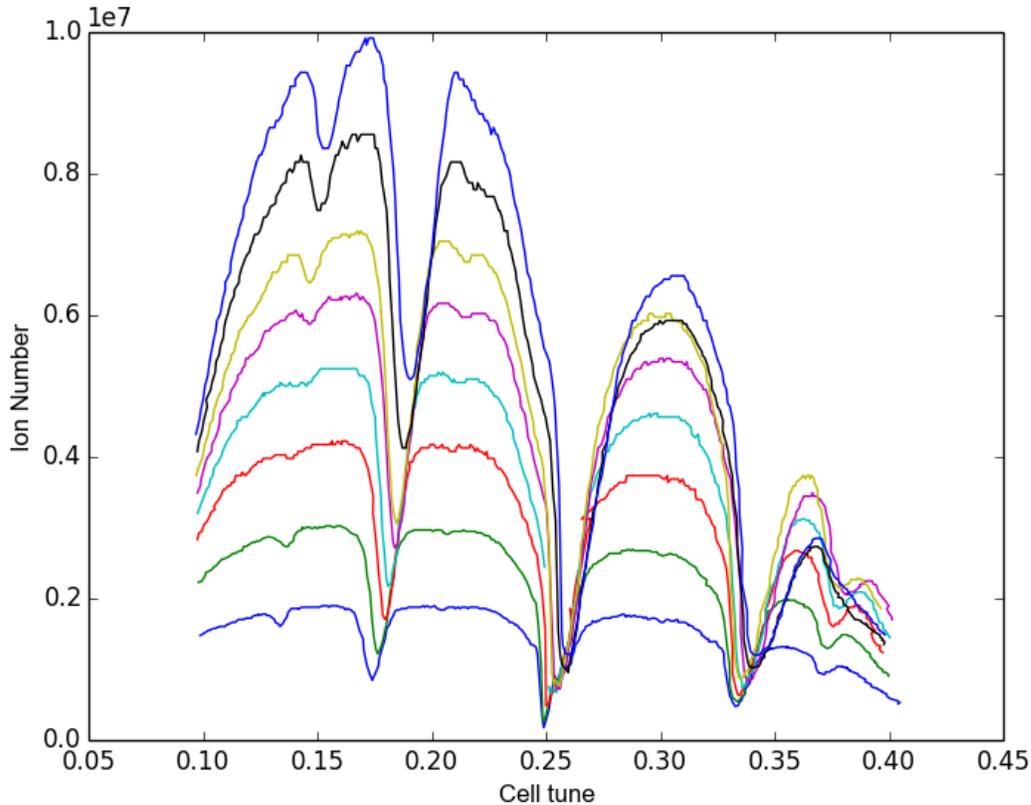}}
\caption{Measured ion number extracted from the S-POD trap against cell tune. Each point on the line represents a separate experiment where the ions are created, stored and extracted.}
\label{fig:Results}
\end{figure*}

\begin{figure}
\makebox[\columnwidth][c]{\includegraphics[width=0.7\columnwidth]{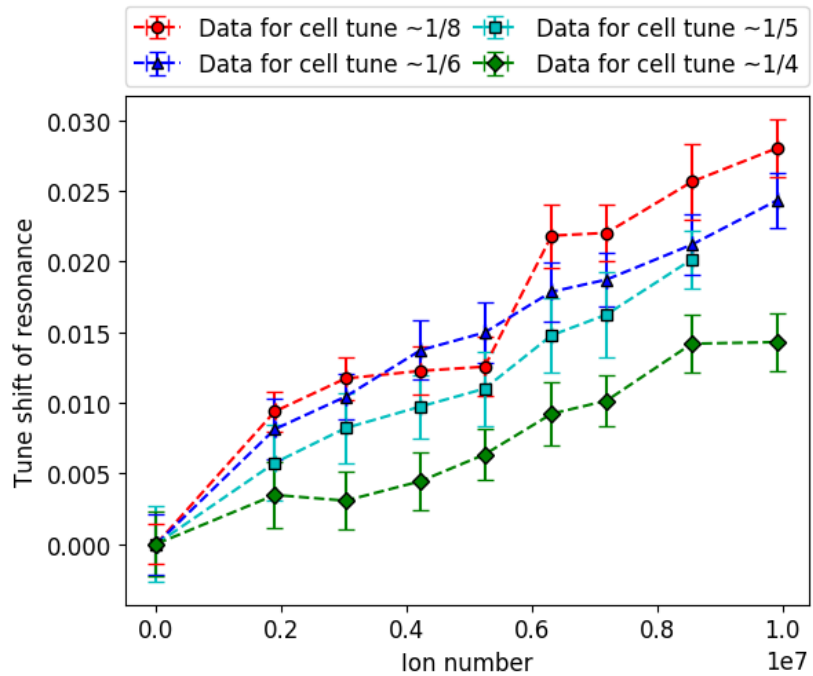}}
\caption{Tune shift against ion number for the four resonances at lowest tune in figure \ref{fig:Results}. Error bars are determined from the width of the resonance minima.}
\label{fig:Analysisplot}
\end{figure}

The experimental data was taken using the S-POD II trap, one of three LPTs at Hiroshima university. To study space charge driven resonances we conducted a number of experiments, each using a different cell tune between the values 0.1 and 0.4. This ensured that a wide range of resonances could be investigated. We then increased the number of ions stored in the trap and repeated the process. As coherent oscillations arise due to space charge forces in high intensity beams large numbers of ions were stored in the trap for these experiments. Measurements were taken for eight different ion numbers between $10^{6}$ and $10^{7}$. 

Identical waveforms were applied to each rod pair (180 degrees out of phase) so that the transverse tunes were always the same.

At each tune value we performed a new experiment with following steps:

\begin{itemize}
\item Argon gas was introduced into the vessel.

\item A sinusoidal voltage was applied to the confining rods with an amplitude that gave a cell tune of 0.15, at this tune the largest possible number of ions are captured by the trap.

\item The electron gun was switched on for less than 1 s to ionise the argon gas. 

\item The captured ions were stored in the trap for 50 $\mu$s to reach equilibrium.

\item The amplitude of the voltage applied to the confining rods was altered smoothly over 100 focusing periods until the tune at which the experiment should take place was reached.

\item At this tune the ions were stored for 100ms, equivalent to $10^{5}$ focusing periods.

\item Ions were extracted by setting to zero the DC voltage on the endcap closest to the MCP.

\item  The signal from the MCP, taken from before the phosphor screen, was recorded to calculate the number of ions extracted.

\end{itemize}

We changed the number of ions stored in the trap and repeated these steps over the same tune range. The number of ions stored in the trap is altered by changing the length of time which the electron gun is on, or by increasing the pressure of argon in the vessel.

In total $\sim$ 330 experiments were performed within the range of tunes, each for a given beam intensity. S-POD is automated and capable of taking one measurement approximately every 10 s and so recording the ion loss over this wide range of tunes took only $\sim$ 1.5 hours.

The relationship between applied voltage and cell tune in the trap is calibrated using data taken at a very low ion number ($\sim 5\times10^{3}$). At this ion number any resonances will not be shifted due to space charge forces. Any offset of the resonances from their expected tune values allows a calibration factor to be calculated. This factor accounts for any systematic errors, such as the actual inscribed radius of the trap differing from 5 mm due to misalignments.

Figure \ref{fig:Results} shows the raw data from the experiments at large ion number, with the calibration applied. 

When the beam has encountered a resonance the resultant beam loss will lead to a reduction in the number of ions extracted. Resonances are represented by local minima in the plot of extracted ion number against tune, as shown in figure \ref{fig:Results}.

Figure \ref{fig:Analysisplot} clearly shows that as the number of particles stored in the trap is increased the location of the resonance shifts to a higher bare tune, as expected due to the increased space charge tune depression. The ion loss observed as the cell tune approaches 0.1 and 0.4 is due to an increase in the maximum beta function at these tunes, reducing the number of ions that can be stored. Furthermore, the depth of the pseudo-potential created by the trap changes with the amplitude of the rf voltage, together, these effects characterise the maximum number of ions that can be stored at a given tune \cite{March2005}.

Ion-neutral collisions are found to lead to beam loss over the 100\,ms time scale of this experiment. However, the resonance is experienced immediately after the tune ramp, just 1\,ms after the ions have been left to stabilise. At this point in time, very few ion-neutral collisions will have occurred. As ion-neutral collisions are a non-conservative process they are not included in the Hamiltonian of the LPT. However, similar collisions will also take place in an accelerator, where Intra Beam Scattering (IBS) can also lead to emittance growth. In this paper we describe the effect of the resonant excitation only.

\section{\label{sec:Simulation}Simulation}
\begin{figure}
\makebox[\columnwidth][c]{\includegraphics[width=0.7\columnwidth]{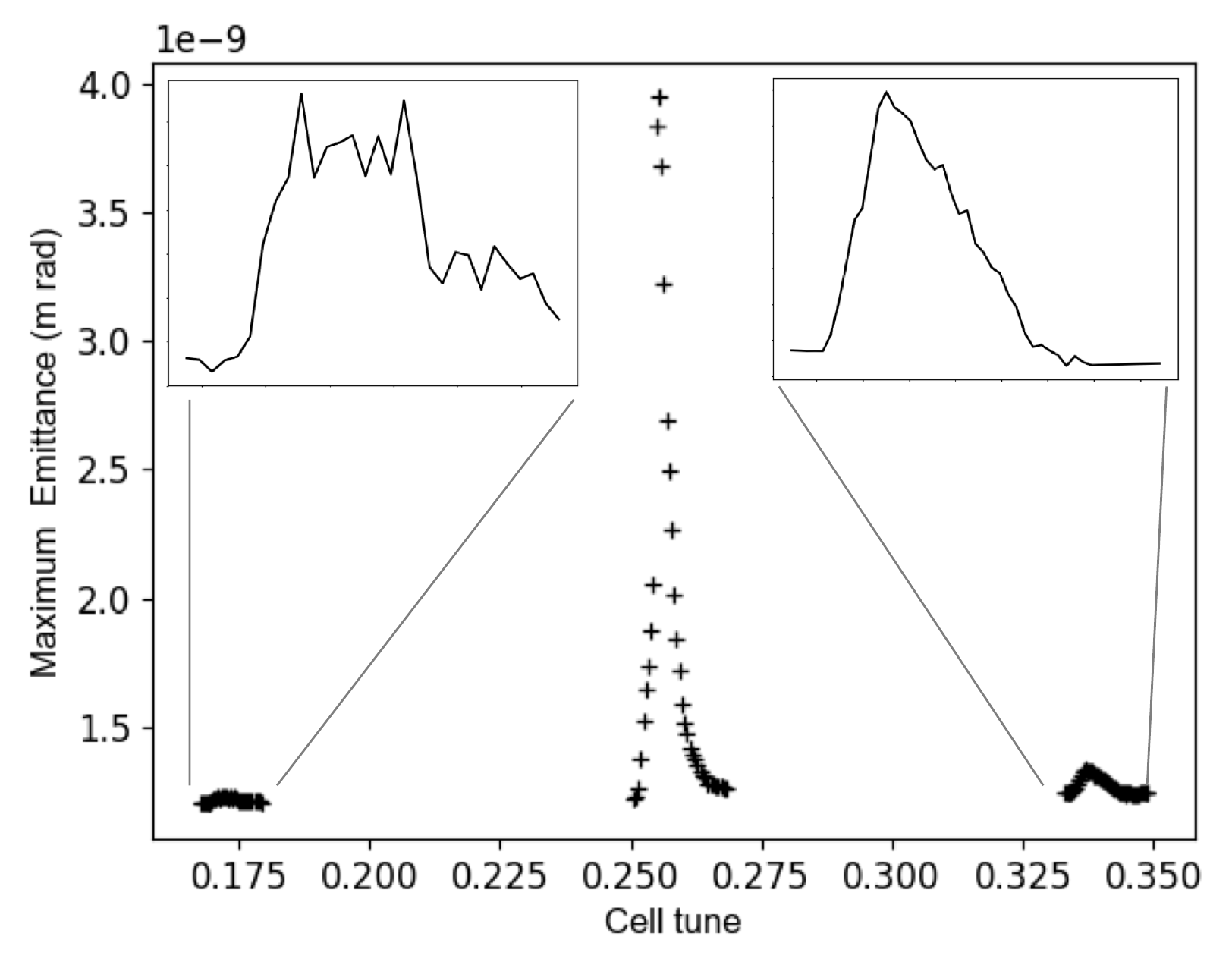}}
\caption{Simulated rms emittance growth at the cell tunes $\frac{1}{6}$, $\frac{1}{4}$ and $\frac{1}{3}$. Left and right insets show enlarged versions of the $\frac{1}{6}$ and $\frac{1}{3}$ resonances respectively.}
\label{fig:emittanceGrowth}
\end{figure}

We use the PIC code Warp \cite{Warp} to simulate the experimental setup in 2D. The simulation applies a voltage to four cylindrical rods, the resultant poisson equation is solved on a grid in the trapping region and when ions are introduced their space charge contribution to the confining potential is included. The simulation is of an ideal LPT, with no misalignments in the trapping rods. Any misalignments in the real system will create additional higher order fields in the trapping region.

The simulation is designed to mimic the experiment as closely as possible, except for the total storage time. A matched distribution of argon ions is injected into the trap at the starting cell tune of 0.15. The ions are trapped at this tune for 50 focusing periods before the tune is ramped to the tune of interest by changing the amplitude of the voltage applied to the confining rods. As in the experiment, this change in tune occurs over 100 focusing periods regardless of the final tune. The ions are then stored in the trap for 550 focusing periods at the final tune, so that the simulation runs for 700 focusing periods in total. 

In the simulation ions are stored for only a fraction of the $10^{5}$ focusing periods used in the experiment. This is due to the significant computational time required to simulate even 550 focusing periods, highlighting the importance of LPTs in understanding resonant effects over longer time scales. 

We used this method to study the increase in the emittance of the distribution at the resonances at cell tunes $\sim \frac{1}{6}$,  $\sim \frac{1}{4}$ and  $\sim \frac{1}{3}$, focusing in greater detail on the resonance at $\sim \frac{1}{4}$. In each simulation the growth in emittance was recorded as well as the evolution of the phase space and any particle loss due to scraping on the rods. Emittance growth is studied instead of particle loss as little, if any, loss occurs on such short timescales. Furthermore, any loss that occurs after only 550 focusing periods is due to the loss of halo ions, and therefore provides very little information on the ion distribution as a whole.

\begin{figure}
\makebox[\columnwidth][c]{\includegraphics[width=0.8\columnwidth]{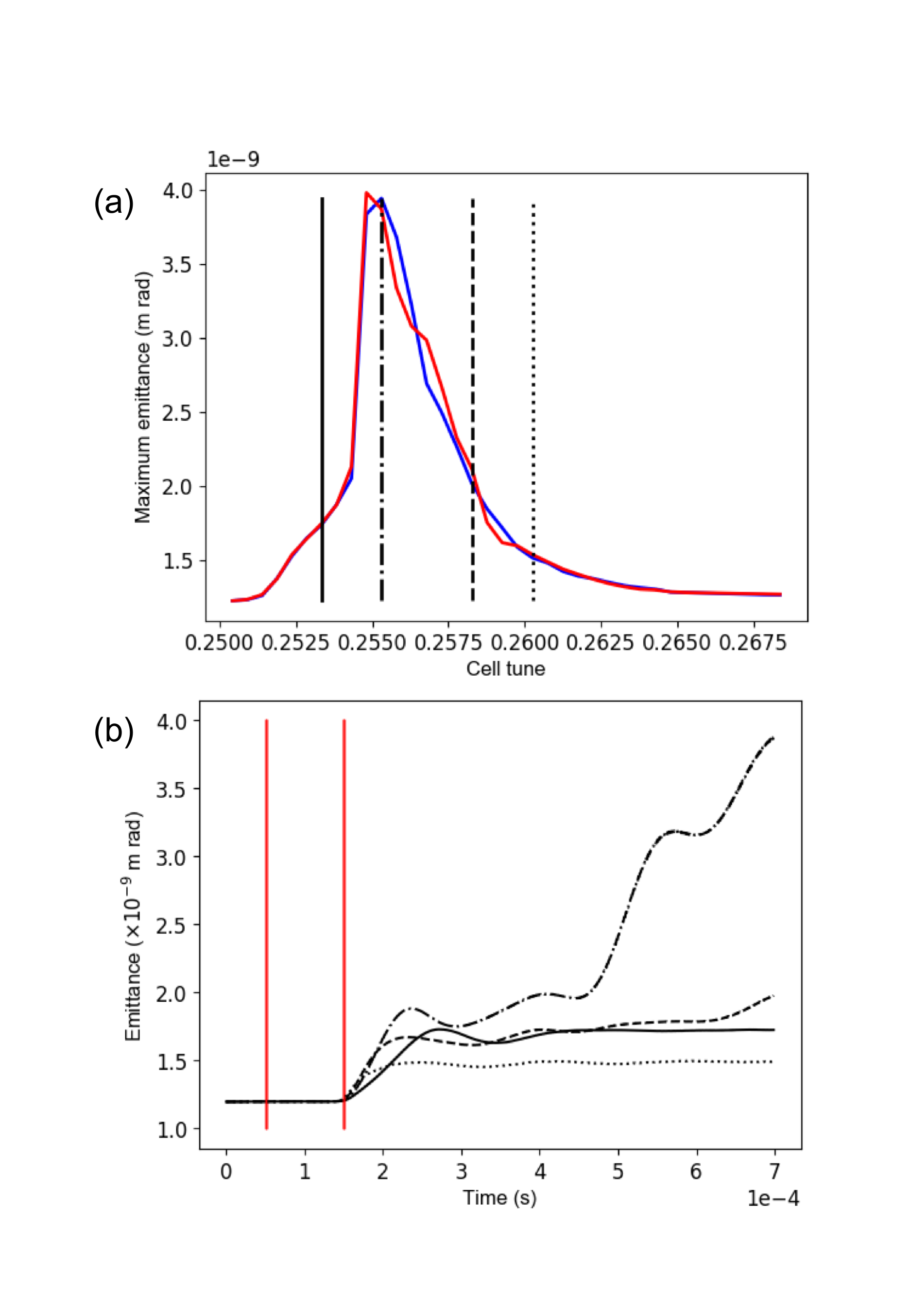}}
\caption{(a) shows the rms emittance growth in the horizontal (blue) and vertical (red) directions at the resonance at cell tune $\frac{1}{4}$, with the four tunes studied in greater detail highlighted by vertical lines. Further information on these tunes is provided in figure \ref{fig:phaseSpace}. (b) shows the rms emittance growth as a function of time for each of the four highlighted tunes, vertical lines represent the start and end of the voltage ramp.}
\label{fig:025Res}
\end{figure}

Figure \ref{fig:emittanceGrowth} shows the rms emittance growth at the 3 resonances for $2\times10^{6}$ trapped argon ions. From the phase space distributions extracted from each simulation we found that even after such a short time the $\frac{1}{4}$ resonance shows signs of coherent motion (see the inset phase space plots in figure \ref{fig:phaseSpace}(b)). However, at cell tunes of $\frac{1}{6}$ and $\frac{1}{3}$ this was not observed. This may be due to the relatively short timescale of the simulation or due to the Landau damping of coherent higher order modes, as predicted by Hofmann in reference \cite{Hofmann2017-2}.

The Warp simulations provided the opportunity to study the effect of the ramp in tune on the distribution of the trapped ions. We found that when trapped ions were ramped to a cell tune of $\frac{1}{4}$ or below the change in emittance during the ramp was negligible. However, when the ions are ramped through the $\frac{1}{4}$ resonance the emittance is increased, the magnitude of this increase depends on the final tune and therefore the rate of resonance crossing. This information was then used to guide the analysis of the data presented in section \ref{sec:Analysis}.

\subsection{The effect of fourth order resonance on the envelope instability}

In the case of the resonance at a cell tune of $\frac{1}{4}$ we chose four tunes across the resonance to study in greater detail. At each of these tunes we tracked the ions in the Warp simulation, and from the positional information we used Numerical Analysis of Fundamental Frequencies (NAFF) \cite{Laskar1992} to extract the tunes of individual macro particles. We also calculated the kurtosis, a measure of the shape of the distribution \cite{Press1988}, at every focusing period. 

We chose to study the resonance at cell tune $\frac{1}{4}$ as the competition between the fourth order incoherent resonance and the second order coherent resonance (also known as the envelope instability) at this tune is the subject of much previous work. Simulations by Hofmann and Boine-Frankenheim in reference \cite{Hofmann2015} show that these two resonances overlap, with the fourth order first exciting particles and the stronger envelope instability taking over as the simulation progresses. In reference \cite{Jeon2009} the fourth order resonance was found to dominate on smaller timescales and its width was measured at GSI. 

\begin{figure}[!htb]
\makebox[\columnwidth][c]{\includegraphics[width=\columnwidth]{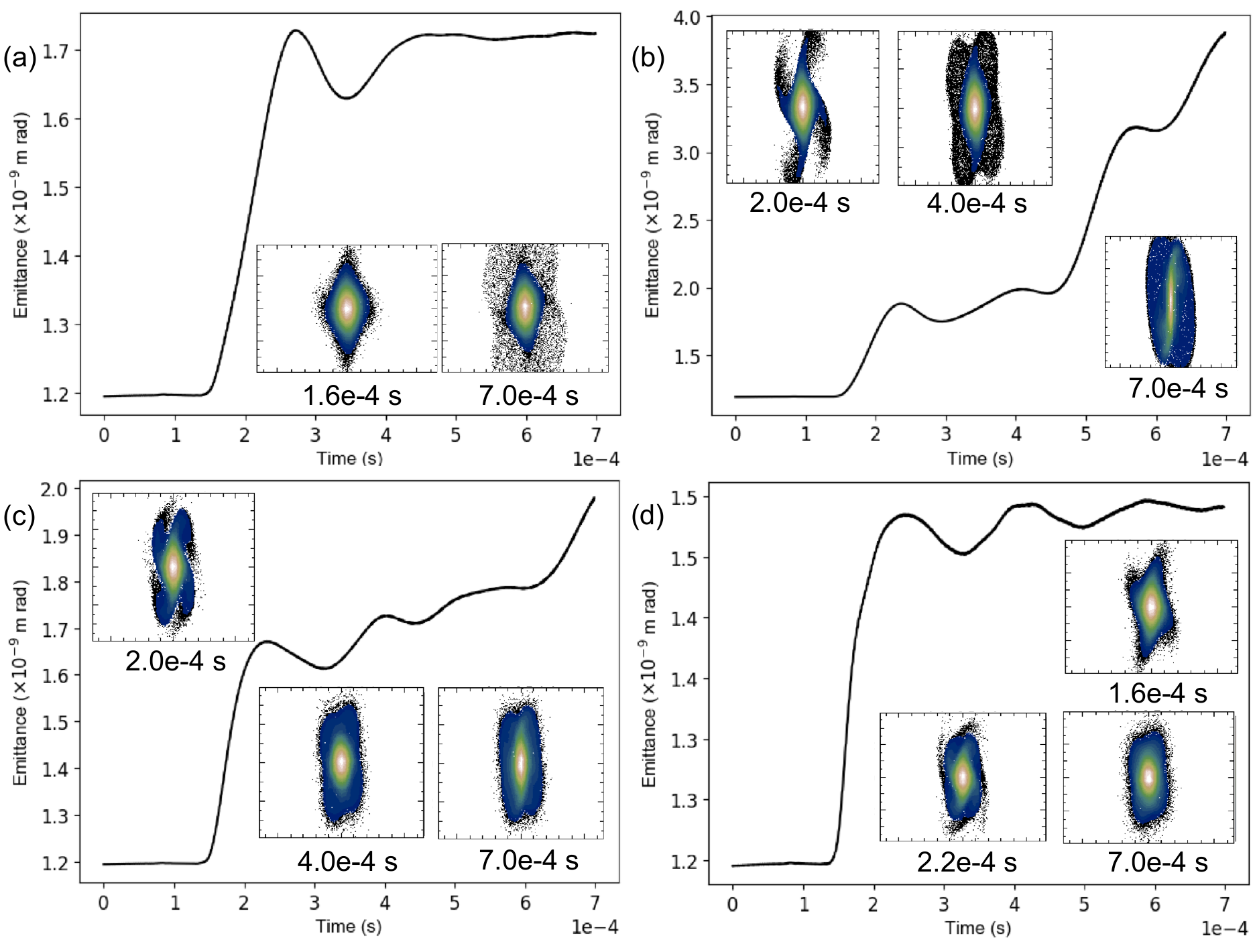}}
\caption{Emittance growth (rms) in the horizontal direction at bare cell tunes (a) 0.253, (b) 0.255, (c) 0.258 and (d) 0.26. Emittance growth in the vertical direction is very similar in each case. Insets show the x x' phase space at points of interest.}
\label{fig:phaseSpace}
\end{figure}

Once the emittance growth due to either the coherent or incoherent resonance is so great that the tune of the beam no longer falls within the stop band of the envelope instability, the distribution can no longer be excited \cite{Li2014}. This leads to the characteristic shape of the coherent resonance, with a sharp increase in particle loss on the lower tune side of the resonance and a gentler gradient on the high tune side. 

On the other hand, the fourth order resonance has the potential to excite particles over much longer time scales, trapping them in the stop band. 

Our simulations agree very well with these previous results. Figure \ref{fig:025Res} shows that the simulated resonance has the characteristic shape of a coherent resonance with the addition of areas of emittance growth outside of this coherent region (for example the region between bare cell tunes of 0.251 and 0.253 in figure \ref{fig:025Res}(a)), where the emittance growth is significantly reduced, but not yet negligible. 

In figure \ref{fig:025Res}(a) a small difference in the emittance growth between the horizontal and vertical directions indicates that the second order resonance dominates at those tunes. The second order resonance is a strong instability and small, random, differences in initial conditions influence in which plane the emittance grows first \cite{Hofmann2017}. In figure \ref{fig:025Res}(a) the four tunes that were studied in greater detail are highlighted with vertical lines and the emittance growth as a function of time at each of these tunes is shown in the lower panel. 

Figure \ref{fig:phaseSpace} shows the evolution of the phase space at the four tunes studied in detail. Together figures \ref{fig:025Res}, \ref{fig:phaseSpace}(a) and \ref{fig:phaseSpace}(d) show that the outer regions of emittance growth (at the higher and lower tunes within the resonance) are due to the excitation of particles through the fourth order resonance, with this excitation at the lower tune side of the resonance leading to the formation of a large halo, which can lead to particle loss over a shorter timescale. In the centre of the resonance the phase space plots in figures \ref{fig:phaseSpace}(b) and \ref{fig:phaseSpace}(c), clearly show a fourth order excitation followed by the stronger envelope (second order) instability, leading to an initial growth in the emittance, a plateau, and then further emittance growth.

\begin{figure}[!htb]
\makebox[\columnwidth][c]{\includegraphics[width=\columnwidth]{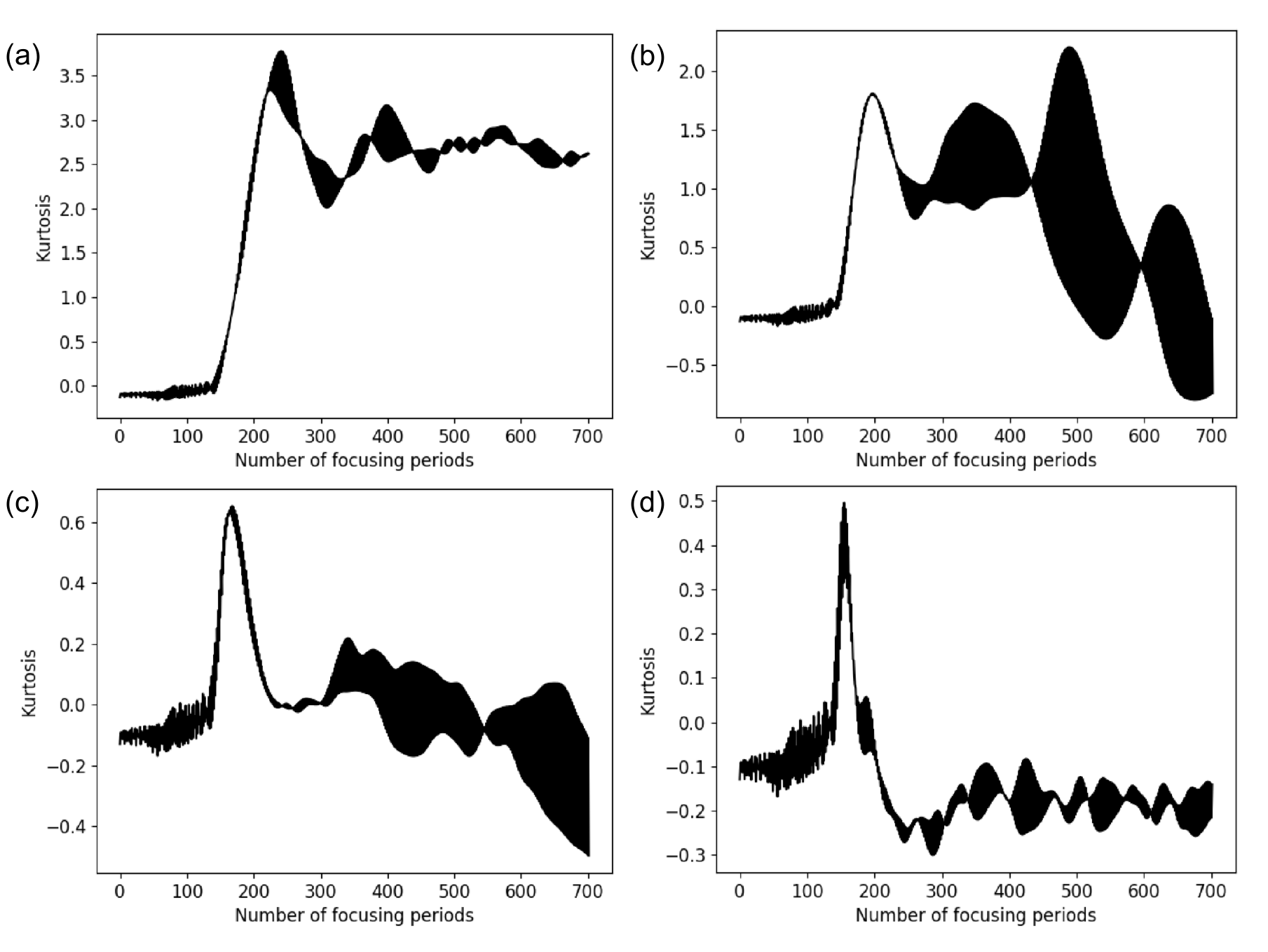}}
\caption{Kurtosis in the horizontal direction at bare cell tunes of (a) 0.253, (b) 0.255, (c) 0.258 and (d) 0.26. Kurtosis in the vertical direction is very similar in each case.}
\label{fig:kurtosis}
\end{figure}

The kurtosis of the distributions is shown in figure \ref{fig:kurtosis}. The kurtosis represents the peakedness of the ion distribution with respect to a Gaussian, which has a kurtosis of 0. Larger positive values indicate a more pointed shape, whereas larger negative values indicate that the distribution is more square. The incoherent fourth order excitation of particle leads to a positive value of the kurtosis as a small number of particles are excited to large amplitudes. In the case of halo formation at a bare cell tune of 0.253 (figure \ref{fig:kurtosis}(a)) the kurtosis remains large, showing that the halo remains an important feature over a large number of cells. In the case of the incoherent excitation at a bare cell tune of 0.26 (figure \ref{fig:kurtosis}(d)) the kurtosis returns to 0, showing that although the emittance increases the distribution is able to redistribute and return to a roughly Gaussian shape. 

In the regions of coherent excitation (figures \ref{fig:kurtosis}(b) and \ref{fig:kurtosis}(c)) the kurtosis first returns to 0 and then becomes negative as the beam is excited as a whole.

\begin{figure*}
\makebox[\columnwidth][c]{\includegraphics[width=0.7\columnwidth]{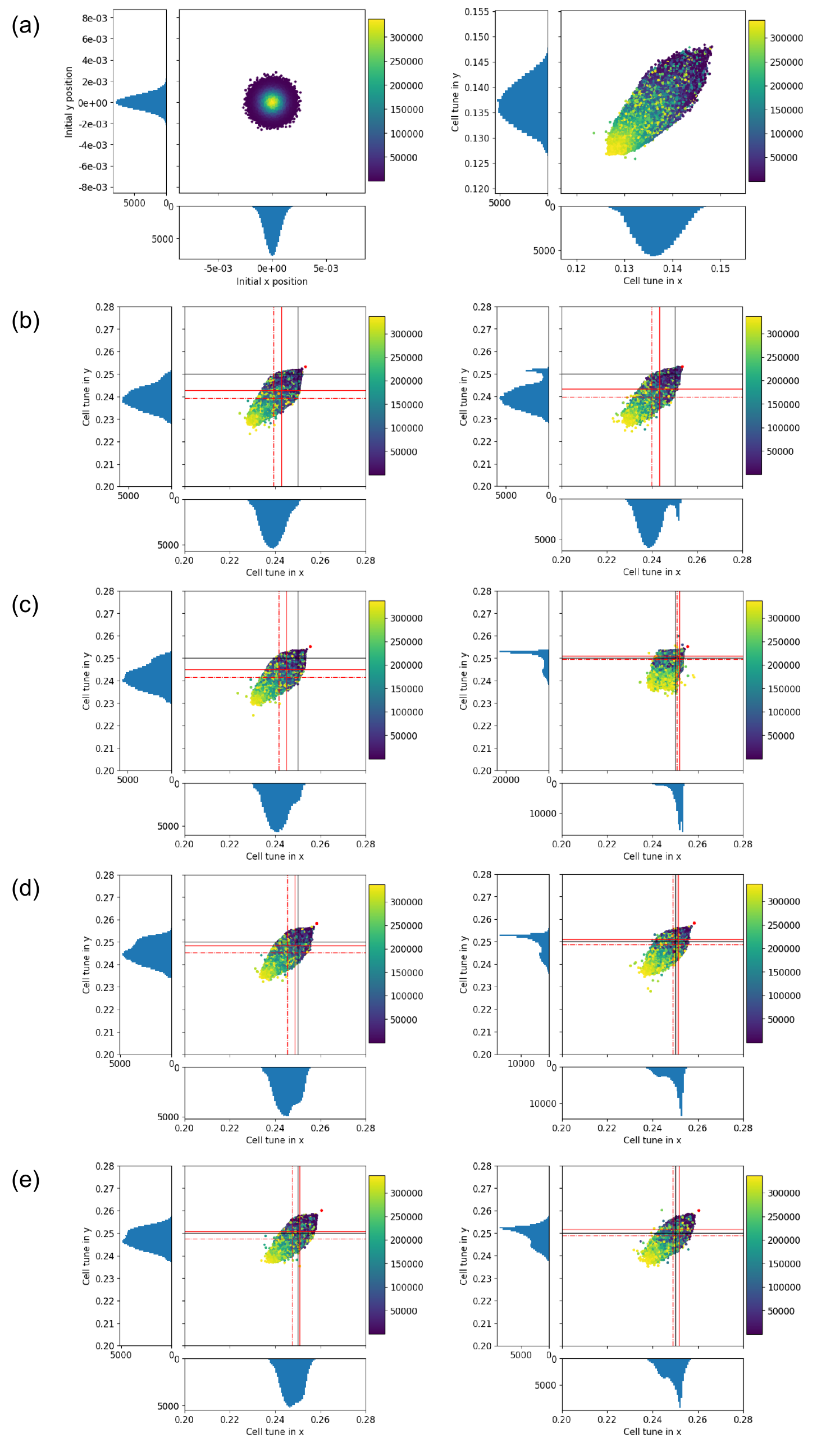}}
\caption{Tune footprints for bare cell tunes of (b) 0.253, (c) 0.255, (d) 0.258 and (e) 0.26. Plot (a) shows the initial distribution in space (left) and the starting tune footprint before the ramp (right). For subplots (b), (c), (d) and (e) the left plot shows the tune footprint immediately after the voltage ramp and the right hand plot shows the tune footprint at the end of the simulation. Distributions are coloured to highlight the relationship between particle location in space and tune. Red dots represent theoretical bare tune based on the applied voltage and solid black lines highlight the location of a tune of $\frac{1}{4}$. Broken red lines show the rms of each tune footprint and solid red lines show the theoretical location of the coherent resonance, assuming $C_{2} = \frac{3}{4}$.}
\label{fig:tune}
\end{figure*}

The tune footprint for the four tunes studied is shown in figure \ref{fig:tune}. The tunes are calculated over 50 focusing periods, firstly at the start of the simulation (at cell tune of 0.15 in each case), then immediately after the tune ramp and finally at the end of the simulation. For each tune footprint the rms tune and the theoretical location of the coherent envelope instability is also plotted.

Figure \ref{fig:tune} shows that the maximum emittance growth does not occur at the theoretical location of the coherent resonance. The maximum emittance growth occurs at a bare cell tune of 0.255 (figure \ref{fig:tune}(c)), instead of the predicted theoretical value of closer to 0.259. This is due to the initial fourth order excitation of the ions, which alters the ion distribution, in turn reducing the tune depression. This, coupled with the finite width of the resonance, means that the coherent resonance does not occur at the tune predicted by the $C_{m}$ factor. At 0.255, fourth order excitation pushes the distribution onto the coherent envelope instability, the ions spend the maximum time within the coherent stopband and therefore the emittance growth is maximised. At 0.258, the distribution is moved towards the exit of the stop band of the envelope instability by the fourth order incoherent excitation, the time spent within the coherent stopband is reduced and so is the emittance growth. 

Through further simulation we confirmed that this remains true even when simulations are extended to a total length of 1350 focusing periods, after which time the coherent emittance growth has plateaued. Simulations which introduce a matched distribution to the trap directly at the tune of interest, without the voltage ramp, verified that this effect is not due to the slow ramp onto the resonance.

Returning now to the experimental work with the S-POD LPT, we therefore expect our experimental results to show evidence of this effect, with the extracted value of $A_{m}$ (equation (\ref{eq:A})) at a cell tune of $\frac{1}{4}$ smaller than $C_{2} = \frac{3}{4}$.

\section{\label{sec:Analysis}Analysis of experimental results and discussion}

When analysing the experimental data we chose to focus on only the four resonances at the lower tune values. These resonances are clearly distinct and it is possible to locate the cell tune where the fewest ions are extracted from the trap for each resonance. Furthermore, the simulations presented in section \ref{sec:Simulation} suggest that at a cell tune higher than $\frac{1}{4}$ the emittance will be increased by passing through the strong envelope instability, complicating the data analysis for these resonances.

We located the point of maximum beam loss for the four resonances studied. The tune corresponding to this point was extracted from the data by a minimum finding algorithm. The change in the tune at which the resonance is located is plotted against the ion number to give figure \ref{fig:Analysisplot}.

\begin{figure}
\makebox[\columnwidth][c]{\includegraphics[width=0.7\columnwidth]{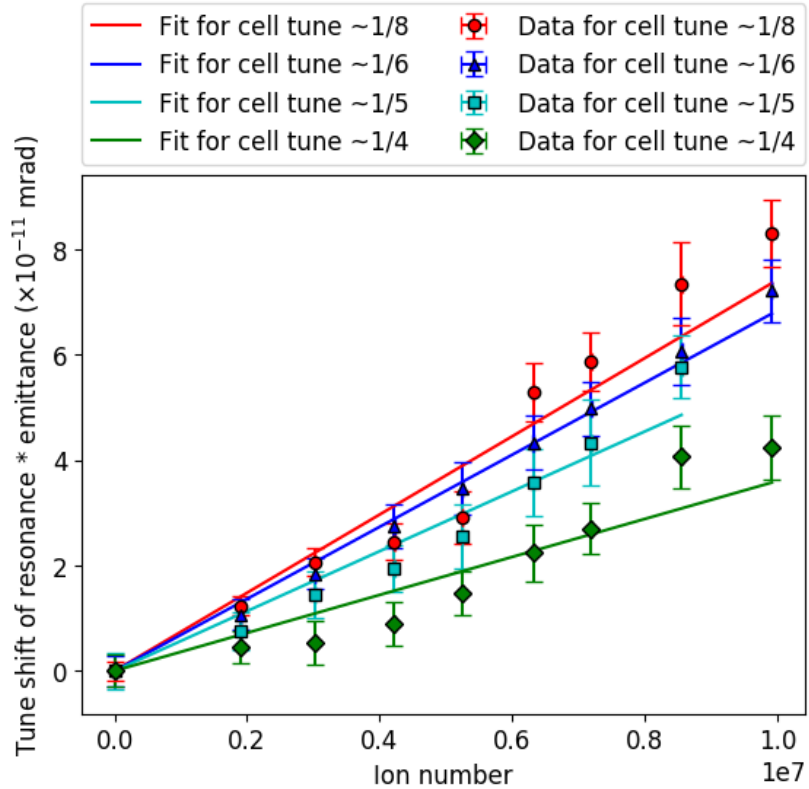}}
\caption{Tune shift multiplied by the ion cloud emittance for the four resonances studied against the ion number. Here error bars include the error on the emittance values and the solid lines show the gradient of a linear fit to the data.}
\label{fig:Analysisplot2}
\end{figure}

\begin{figure}
\makebox[\columnwidth][c]{\includegraphics[width=0.7\columnwidth]{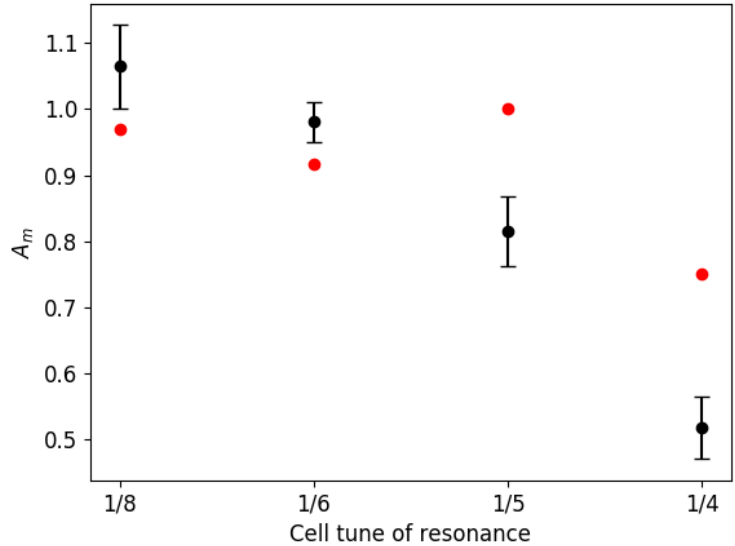}}
\caption{Values of $A _{m}$ from the fitting process in figure \ref{fig:Analysisplot2} with error bars based on fit in figure \ref{fig:Analysisplot2} and the calculation of the ion cloud emittance at each ion number. The red points represent the theoretical $C_{m}$ values for the lowest order space charge driven resonance at each cell tune.}
\label{fig:cm}
\end{figure}

A number of steps are required to extract $A_{m}$ (from equation (\ref{eq:A})) from the data in figure \ref{fig:Analysisplot}. Firstly the expected single particle tune shift at each intensity should be calculated. To do this a value of the emittance of the ion cloud and the longitudinal length of the distribution in the trap are required. Then, a linear fit can be applied to a plot of the tune shift of each resonance multiplied by the emittance, against the ion number in the trap (figure \ref{fig:Analysisplot2}). Any deviation from the expected single particle tune shift indicates that $A_{m}$ deviates from 1. Final values for $A_{m}$ are presented in figure \ref{fig:cm}.

To calculate the expected single particle tune shift, $\Delta Q$, the ions are assumed to occupy a cylinder with a transverse Gaussian distribution and length $L$ \cite{Schindl}. As ion motion in the Paul trap is non-relativistic magnetic forces can be neglected and the rms tune shift of the equivalent beam is calculated to be 

\begin{equation}
\label{eq:deltaQ}
\Delta Q = \frac{\lambda_{\rm rf}r_{\rm p}}{8 \pi}\frac{N}{L \epsilon_{\rm a}},
\end{equation}

\noindent
where $N$ is the number of ions stored in the trap, $\epsilon_{\rm a}$ is the rms emittance, $r_{\rm p}$ is the classical particle radius and $\lambda_{\rm rf}$ the wavelength of the applied rf voltage. 

We assume a Gaussian distribution based on significant evidence from previous S-POD experiments in which the MCP is used to image the ion distribution \cite{Moriya2015, Ito2009}. Calculating $\Delta Q$ from equation (\ref{eq:deltaQ}) gives a result that agrees very well with the rms tune shift calculated from the Warp simulations through NAFF. 

The ion distribution length $L$ was determined experimentally using IBEX. As both traps are designed to be longitudinally identical and the same electron gun is used, the result applies to both S-POD and IBEX (see figure \ref{fig:SPODSetup}). Further details of this experiment can be found in \ref{Effective plasma length measurement}. 

Using the equations presented in section \ref{sec:Emittance and Temperature in a linear Paul trap} and the experimentally determined value for the longitudinal distribution length, the emittance was calculated for each ion number. A single emittance value cannot be assumed as the increased space charge forces at higher intensity will distort the potential experienced by the ions. The tune depression for the resonance at a cell tune of $ \frac{1}{6}$ was used to calculate the rms tune depression, at first assuming $A_{m} = 1$. This was then used to calculate the transverse temperature. The temperature was used to find a value for the rms beam size and finally the rms emittance. 

The data in figure \ref{fig:Analysisplot} was combined with the results of the emittance calculation at each ion number and a linear fit was applied, as shown in figure \ref{fig:Analysisplot2}. The gradient of the fit gives $A_{m}$ for the resonance, as 

\begin{equation}
\label{eq:deltaQ2}
\frac{\epsilon_{\rm a} \Delta Q }{N} = \frac{ \lambda_{\rm rf} r_{\rm p }}{8 \pi} \frac{A_{m}}{ L }.
\end{equation}

This process was repeated iteratively for the $\frac{1}{6}$ resonance until the factor of $A_{m}$ included in the calculation of the rms tune depression equaled the factor extracted from the fit. The emittance values for this case are presented in Table \ref{table:emittance}.

\begin{table} 
\caption{\label{table:emittance}Table showing the transverse temperature, rms beam size and rms emittance calculated for each ion number studied. }
\begin{indented}
\lineup
\item[]\begin{tabular}{@{}*{4}{l}}
\br
Ion number & Transverse temperature  & rms beam size   &  rms emittance \\
 ($*10^{6}$) &  (eV)  & (mm)   & ($*10^{-3}$ mm mrad) \\
\mr
1.897 & 0.144 $\pm$ 0.003 & 0.670 $\pm$ 0.008 &  1.318 $\pm$ 0.030\\
3.026 & 0.190 $\pm$ 0.004 & 0.779 $\pm$ 0.008 &  1.760 $\pm$ 0.036\\
4.223 & 0.211 $\pm$ 0.005 & 0.837 $\pm$ 0.009 &  1.998 $\pm$ 0.045\\
5.245 & 0.244 $\pm$ 0.006 & 0.906 $\pm$ 0.010 &  2.320 $\pm$ 0.052\\
6.316 & 0.252 $\pm$ 0.006 & 0.934 $\pm$ 0.010 &  2.430 $\pm$ 0.054\\
7.192 & 0.275 $\pm$ 0.006 & 0.981 $\pm$ 0.010 &  2.664 $\pm$ 0.056\\
8.554 & 0.291 $\pm$ 0.007 & 1.024 $\pm$ 0.011 &  2.886 $\pm$ 0.064\\
9.917 & 0.297 $\pm$ 0.006 & 1.051 $\pm$ 0.011 &  2.967 $\pm$ 0.063\\
\br 
\end{tabular}
\end{indented}
\end{table}

This analysis relies on the further assumption that the tune change from a cell tune of 0.15 (where the ions are collected) to the operating point, is adiabatic and that during this process no ions are lost. This assumption is well supported by the Warp simulations.

In the experimental data in figure \ref{fig:Results} the resonance at a cell tune of $\frac{1}{4}$ shows the characteristic shape of a coherent resonance (i.e. the dip in ion number in tune space is asymmetric, with a much steeper slope for tunes lower than the point where the ion number reaches a minimum), with the addition of ion loss at higher and lower tunes. This additional ion loss is explained by the Warp simulation as due to the excitation of a subset of ions through the fourth order resonance. The shape of the experimental ion loss data agrees well with the emittance growth in the Warp simulation, supporting the idea that emittance growth is a good predictor for ion loss. This also suggests that a simulation of only 1350 focusing periods may be sufficient to predict ion loss over longer time scales. The resonance shows a similar shape as intensity is increased.

The other resonances studied experimentally ($\frac{1}{8}$, $\frac{1}{6}$ and $\frac{1}{5}$) do not have the sharp increase in ion loss that usually indicates a coherent resonance, suggesting that ion loss may not be due to coherent oscillations. 

The fitting process is performed for all four resonances and the results of this fit are shown in figure \ref{fig:cm}, alongside the $C_{m}$ factors for purely coherent resonances.

The values of $A_{m}$ extracted from the resonances at tunes of $\frac{1}{8}$ and $\frac{1}{6}$ are consistent with 1, rather than the $C_{m}$ factor for the coherent resonance which is expected to occur at that tune (also plotted in figure \ref{fig:cm}), again suggesting that these resonances may not be coherent.  As it cannot be claimed that the excitation of a subset of particles is simply an incoherent effect, as the motion of these particles effects the rest of the beam through the Coulomb interaction, it is interesting to note that the peak ion loss still occurs at roughly the tune with the highest particle density. 

A value of $A_{m} = 0.517 \pm 0.091$ is extracted for the $\frac{1}{4}$ resonance. This is smaller than the other resonances and indeed smaller than the analytically predicted value of $\frac{3}{4}$. This agrees with the Warp simulations, we expect to see a greater coherent advantage due to the fourth order resonance first increasing the emittance of the beam. 

Any resonances present in a LPT are clearly complex and the product of competing sources of perturbation, as in an accelerator. Whether these resonances are coherent or incoherent, the values of $A_{m}$ presented in figure \ref{fig:cm} show the relative location of the beam loss for a resonance of a given order in a Gaussian beam. These factors help us understand the real behaviour of an accelerator and are directly relevant in future accelerator design. 

Further work is possible to build on this study. The stronger resonances at cell tunes of $\sim\frac{1}{4}, \sim\frac{1}{6}$ and $\sim\frac{1}{8}$ can be studied with a lower storage time to verify that the values of $A_{m}$ remain unchanged. There is also the potential to study anisotropic beams using a LPT, as non-circular beams are expected to have different $C_{m}$ factors \cite{Hofmann1998}. 

Further simulation work to study the $\frac{1}{4}$ resonance at increasing intensities is also desirable to determine the effect of intensity on the relationship between the fourth order resonance and the envelope instability. 
 
To understand fully the mechanism behind beam loss at these resonances further diagnostics are required in the experimental setup. Detection of the coherent oscillations via the induced current on the confining rods should be attempted. Further diagnostics would also allow for an emittance measurement to be made at each ion number. This would improve the accuracy of the extracted values of $A_{m}$.

\section{\label{sec:Conclusion}Conclusion}

This paper presents the first quantitative study of the interaction and difference between coherent and incoherent resonances in a Paul trap, and furthermore uses simulation to explain why the experimental results differ from those theoretically predicted. A LPT has previously been used to study the behaviour of a beam only qualitatively, showing the shift in resonance location with beam intensity. Here we extract numerical values describing the locations of resonances at high intensity.

We have shown through simulation using the PIC code Warp that in a LPT, as in an accelerator, we expect to see coherent oscillation at a cell tune of $\frac{1}{4}$. Here the coherent resonance is in competition with the fourth order resonance, which has a wider stopband. For $2\times10^{6}$ ions, simulations show that, even at tunes where coherent oscillations occur, ions are first excited by the incoherent fourth order resonance. By analysing the tunes of trapped  ions through NAFF the tune footprint at a number of tunes across the resonance was extracted. This showed that the peak in emittance growth did not occur at the location predicted by the $C_{m}$ factor for coherent resonances. We identified that this was due to both the finite width of the envelope instability and the fourth order excitation. 

The shape of the ion loss at the $\frac{1}{4}$ resonance agrees well with shape of increased emittance growth in the Warp simulation, suggesting that a short simulation may be in some cases sufficient to predict beam loss over longer timescales. The experimental results show that the $C_{m}$ factor for a second order coherent resonance does not accurately predict the location of the resonance. Instead we find that the factor describing the shift of the resonance with intensity is less than the theoretical $C_{2}=\frac{3}{4}$, due to the interaction between the fourth order incoherent resonance and the second order coherent resonance at this tune, agreeing with Warp simulation predictions. 

Simulation over 700 focusing periods did not show any signs of coherent oscillation for the resonances at cell tunes $\frac{1}{6}$ and $\frac{1}{3}$. 

The experimental ion loss for the resonances at cell tunes $\frac{1}{8}$, $\frac{1}{6}$ and $\frac{1}{5}$ did not show the shape of coherent resonance and the location of the peak in ion loss occurred in a location consistent with the rms of the tune distribution, suggesting that at this tune coherent oscillation may indeed be Landau damped for a Gaussian beam.

\ack
The Authors would like to thank the beam physics group at Hiroshima university for the experimental time on the SPOD experiment and for their advice and comments. 

We would like to thank Dr Kiersten Ruisard for her advice on the use of Warp.

S.L. Sheehy gratefully acknowledges the support of The Royal Society. 


\section*{References}
\bibliography{References}

\appendix
\section{Longitudinal ion distribution length measurement\label{Effective plasma length measurement}}

To measure the length of the ion distribution in S-POD, the IBEX trap was used (for reasons of experimental availability).

The trap was set up to operate as described in section \ref{sec:Experimental procedure and results}. The ions were extracted from the Faraday cup end of the trap by lowering the endcap voltage to a non-zero DC offset. This allows only those ions with kinetic energy above the DC offset to escape the potential well and reach the faraday cup.

We systematically lowered the endcap voltage and recorded the number of ions extracted in each experiment. The data were fitted to assuming a Maxwell-Boltzmann temperature distribution. From the fit the temperature of the ions was calculated to be 0.378 $\pm$ 0.112 eV. Once the temperature was known the  length was calculated from the knowledge of the potential well created by the endcaps. For this temperature charge fills a cylinder with effective length 66.19 $\pm$ 1.46 mm. 

This low longitudinal temperature results in longitudinal motion over much greater timescales than the transverse oscillations, allowing any effect of the longitudinal motion on the transverse dynamics to be ignored.

\end{document}